\documentclass[aps,pra,superscriptaddress,twocolumn]{revtex4-1}
\usepackage{color,graphicx,hyperref,ulem}
\usepackage{amsmath, amssymb}

\definecolor{green}{rgb}{0,0.7,0.1}
\definecolor{blue}{rgb}{0,0.1,0.7}
\definecolor{red}{rgb}{0.8,0,0}
\definecolor{violet}{rgb}{0.658, 0.0, 0.8312}
\definecolor{blueish}{rgb}{0, 0.3, 0.7}
\definecolor{orange}{rgb}{1, 0.4, 0}
\definecolor{gray}{rgb}{0.2, 0.2, 0.2}

\begin{document}

\title{Quantum effects of operator time ordering in the nonlinear Jaynes-Cummings model}

\author{Tobias Lipfert}\email[]{tobias.lipfert@univ-lille.fr}
\affiliation{Univ. Lille, CNRS, UMR 8523 - PhLAM - 
	Physique des Lasers Atomes et Mol\'ecules, F-59000 Lille, France}

\author{Fabian Krumm}\email[]{fabian.krumm@uni-rostock.de}\affiliation{Arbeitsgruppe Theoretische Quantenoptik, 
Institut f\"ur Physik, Universit\"at Rostock, D-18059 Rostock, Germany}

\author{Mikhail~I.~Kolobov}
\affiliation{Univ. Lille, CNRS, UMR 8523 - PhLAM - Physique des Lasers Atomes et Mol\'ecules, F-59000 Lille, France}

\author{Werner Vogel}\affiliation{Arbeitsgruppe Theoretische Quantenoptik, Institut f\"ur Physik, Universit\"at Rostock, D-18059 Rostock, Germany}

\date{\today}

\begin{abstract}
		Recently, in [Phys. Rev. A \textbf{97}, 043806 (2018)], the detuned and nonlinear Jaynes-Cummings model describing the quantized motion of a trapped ion was introduced and its corresponding dynamics was solved via considering the driving laser in a quantized manner.
		In this work we reconsider this model and show that it can likewise be solved with a classical driving laser field.
		Using the exact solution we investigate the quantum time-ordering effects of the system with respect to nonclassicality of the motional states of the ion.
		Furthermore, we use the Magnus expansion to analyze the impact of certain orders of the time ordering and derive and exact radius of convergence beyond the established and only sufficient criterion.
		Finally, the differences of the solution derived here and the previously found one using a quantized pump, are discussed.
\end{abstract}

\maketitle

\section{Introduction }

	When one quantizes the electromagnetic field, classical field variables are replaced by field operators---see for example~\cite{Schleich,VogelW2006,GrynbergAF2010,Agarwal2013}.
	The quantum nature of light is reflected in the non-commutative algebra of these operators which in turn give rise to non-zero non-equal-time commutators of certain observables.
	Thus, temporal correlations may arise that are not covered by the classical theory of Maxwell.
	One phenomenon of this category is the quantum effect of photon antibunching~\cite{Walls76,Mandel76}, whose experimental verification in 1977~\cite{KDM77} may be considered as the final proof to Einsteins light quanta Hypothesis~\cite{Einstein}.
	A more general discussion of the underlying field inequalities is among others given in Refs.~\cite{M99,V08}.
	
	The evolution of dynamical systems with time-dependent perturbations are the subject of time-dependent perturbation theory, cf.~\cite{LandauL1977,Griffiths1995,Fliessbach2008}.
	Such systems are usually treated in terms of the Dyson series~\cite{Dyson1949} and approximations based on the latter.
	In quantum optics non-equal-time commutators give rise to quantum time-ordering effects~\cite{KVW87}.
	In Ref.~\cite{ChristBMS2013} the latter were studied for the processes of sum frequency generation and parametric down conversion.
	Recently, in the study of quantum time-ordering effects in dynamical systems the Magnus expansion~\cite{Magnus1954,BlanesaCOR2009} has been considered as a useful representation of the corresponding evolution operators, cf.~\cite{QuesadaS2014,QuesadaS2015,QuesadaS2016,KrummSV2016,KrummV2018,LipfertHPK2018}.
	Let us note that, since noncommutativity of quantum-mechanical operators is a pure quantum effect, it can be used for quantitative measure of nonclassicality of a quantum state, as very recently proposed in Ref.~\cite{BievreHPK2018}.
		
		Remarkably, the Magnus expansion allows for the formulation of approximations in terms of different orders of nested non-equal-time commutators.
		As such approximations remain within the Lie algebra of whatever space is spanned by the non-equal time commutators, important symmetries of the studied systems usually remain preserved~\cite{BlanesaCOR2009,LipfertHPK2018,QuesadaS2014,QuesadaS2015,QuesadaS2016}.
		Furthermore, the first-order approximation corresponds exactly to the case of neglecting ordering effects. 
		This allows for a clear identification of ordering effects.
		
		When approximations of the time-evolution operator are formulated in terms of the Magnus expansion, an increasing number of Magnus orders leads to a stepwise inclusion of time-ordering effects.
		That is, with higher orders  one moves closer to the correct dynamics of the system---i.e., the incorporation of all time-ordering effects.
		However, this is limited by two factors, 
		(i) the expressions for higher-order corrections can take quite complex forms, cf.~\cite{PratoL1997}, which may make their evaluation quite tedious and 
		(ii) the Magnus expansion generally only works withing a finite radius of convergence, that is it may diverge at some point and the correct dynamics of the system cannot be recovered in terms of increasing orders of corrections.
		
		In  case of divergence a  comparison of the cases of neglected time ordering with the case of time ordering in terms of the Magnus expansion will lead to misinterpretations.
		Admittedly, for small time scales the Magnus expansion does always converge but significant deviations caused by the negligence of ordering effects may only arise after sufficiently long times.
		Thus, precise knowledge of the limits of convergence is in our case indispensable.
		Indeed, there exist sufficient upper bounds for evolution periods where convergence occurs~\cite{BlanesaCOR2009}, but exact upper bounds can generally only be found for generic cases.
		
		Interestingly, for the detuned nonlinear Jaynes-Cummings dynamics of a trapped ion as studied in Ref.~\cite{KrummV2018} an analytical solution of the dynamics could be found in the case of a quantized pump.
		With this ansatz, the system became time independent in the Schr\"odinger picture, i.e., no ordering effects occurred.
		However, this of course also prevents a discussion in terms of time-ordering effects.
		In the context of time-ordering effects in this system, a solution of the dynamics without the detour of pump quantization seems desirable.
		This is exactly what we present here and, furthermore, we demonstrate how this approach can be used to derive an exact radius of convergence of the Magnus expansion.

	The paper is structured as follows.
	At first we shortly reconsider the detuned nonlinear Jaynes-Cummings model and present the theoretical background in Sec.~\ref{Sec:model}.
	After that, an analytic solution for its dynamics is derived in Sec~\ref{Sec:exactevolution2} which we use to discuss the influence of time ordering on nonclassicality in Sec.~\ref{Sec:timeOrderingNCL}.
	Then, in Sec.~\ref{Sec:TOcorrections}, we further investigate the impact of certain orders of time-ordering corrections and especially analyze the convergence of the approximation.
	Afterwards, in Sec.~\ref {Sec:NCLpump}, we compare the solutions found in this work with the one derived in~\cite{KrummV2018}.
	  Finally a summary and some conclusions are given in Sec.~\ref{Sec:conclusion}.

\section{Model }
\label{Sec:model}

	Let us briefly recapitulate the physical model to be studied and present the theoretical background.
	If an ion is caught in a Paul trap, its motion can be described in a quantized manner, see Refs.~\cite{B92a,B92b,CBZ94} or Chap. 13 of~\cite{VogelW2006}.
	The resulting states of the ion are referred to as motional or vibrational states.
	Via the interaction of the ion with optical radiation, e.g. a laser, the generation of a plethora of motional states became feasible~\cite{C93,M96,F96,MV96,GS96a,GS96b,G97,MM96,G96,WV97}.

\subsection{Detuned nonlinear Jaynes-Cummings Hamiltonian}
\begin{figure}[th]
	\centering
		\includegraphics*[width=8.6cm]{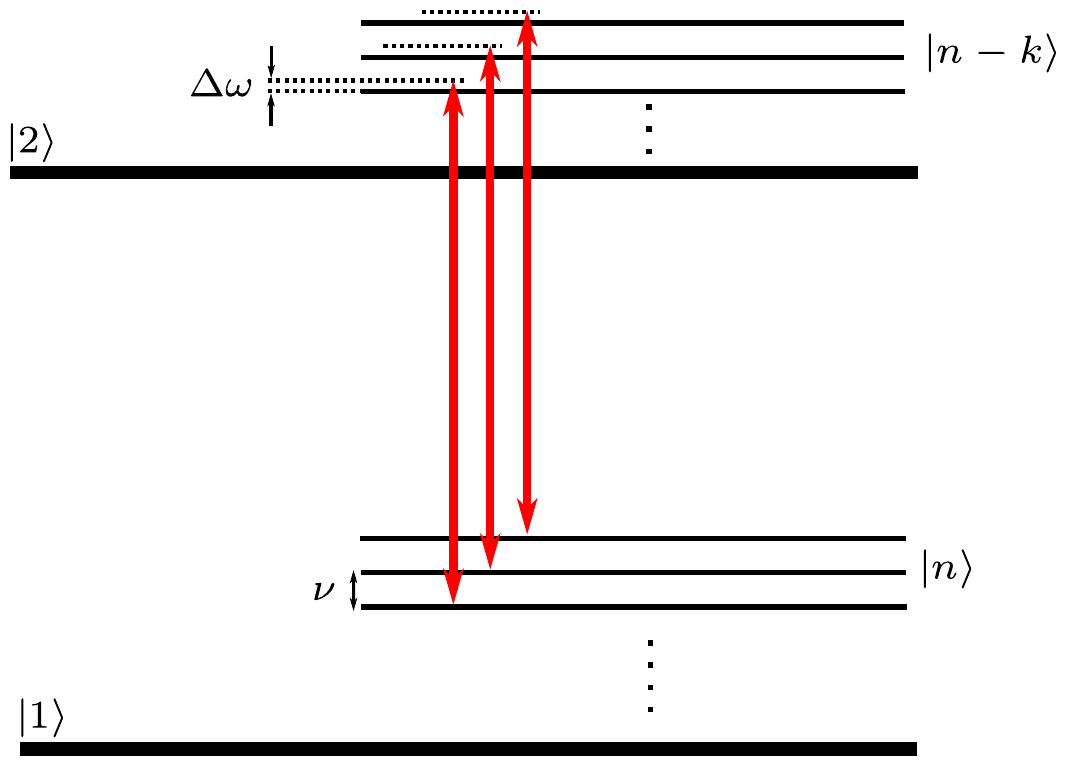}
		\caption{
			Scheme of the physical system described by the interaction Hamiltonian in Eq.~\eqref{Eq:Hinteraction}. 
			The electronic ground state, denoted with $|1\rangle$, and the corresponding excited state $|2 \rangle$, are separated by the electronic transition frequency $\omega_{21}=\omega_2-\omega_1$. 
			In a harmonic trap potential the vibrational levels are equidistantly separated by the trap frequency $\nu$.
			The frequency of the driving laser (red arrows)  is detuned from the $k$-th sideband by $\Delta \omega$.
			That is, $\omega_L=\omega_{21} - k \nu + \Delta \omega$.
			}\label{Fig:scheme}
	\end{figure}

	The full dynamics describing the interaction of an ion with a laser is rather complicated and can, in general, only be solved numerically.
	However, under certain but realistic approximations the interaction Hamiltonian of the system can be simplified to a nonlinear generalization of the Jaynes-Cummings Hamiltonian~\cite{V95}, describing the electronic coupling to the $k$-th sideband.
	In the interaction picture, including a detuning $\Delta \omega$, it reads
	 \begin{align}
	 	  \label{Eq:Hinteraction}
		\hat H_\text{int}(t)= \hbar |\kappa| e^{- i\Delta \omega t + i \theta} \hat A_{21}  \hat f_k(\hat a^\dag \hat a;\eta) \hat a^k+ \text{H.c.} 
	 \end{align}
	A more detailed derivation of the Hamiltonian is given in Ref.~\cite{KrummV2018}.
	The corresponding scheme is depicted in Fig.~\ref{Fig:scheme}. 
	
	In Eq.~\eqref{Eq:Hinteraction} the following quantities and definitions are used:
	$\kappa=|\kappa|e^{i \theta}$ describes the strength of the coupling of the electronic to the vibrational levels of the ion (vibronic coupling), where $|\kappa|$ growth linearly with the pump amplitude.
	$\hat A_{ij}=|i\rangle \langle j|$ ($i,j=1,2$) are the atomic flip operators corresponding to the $|j\rangle \rightarrow |i\rangle$ transitions.
	The operator function $\hat f_k(\hat a^\dag \hat a;\eta)$ describes the mode structure of the laser field in the case of a standing wave at the (operator-valued) position of the ion.
	It reads as,
	\begin{align}
		 \hat f_k(\hat a^\dag \hat a;\eta)= \frac{1}{2} e^{i \Delta \phi -\eta^2/2} \sum_{l=0}^\infty \frac{(i \eta)^{2l+k}}{l! (l+k)!} \hat a^{\dag l} \hat a^l &+ \text{H.c.}  		
	\notag\\
		=\frac{1}{2} e^{i \Delta \phi -\eta^2/2}  \sum_{n=0}^\infty |n\rangle \langle n | \frac{(i \eta)^k n!}{(n+k)!} L_n^{(k)} (\eta^2) &+ \text{H.c.}  , 
		\label{Eq:operatorfuncf}
	\end{align}
	with $L_n^{(k)} $ denoting the generalized Laguerre polynomials.
	That is, via $ \hat f_k(\hat a^\dag \hat a;\eta)$ a nonlinear dependence of the dynamics on the excitation of the vibrational mode is obtained.
	The quantity $\Delta \phi$ is the relative position of the trap potential to the laser wave and $\eta$ is the Lamb-Dicke parameter.
	
	The operators $\hat a^\dag$  ($\hat a$) are the creation (annihilation) operators of the motional excitation with frequency $\nu$.
	The latter is contained in the Hamiltonian of the free motion,
	\begin{align}
		\hat H_0 = \hbar \nu \hat a^\dag \hat a + \hbar \omega_{21} \hat A_{22},
	\end{align}
	$\omega_{21}=\omega_2 - \omega_1$ is the separation of the electronic levels $|1\rangle$ and $|2 \rangle$.

	 We are interested in the situation where the classically described laser, with the frequency $\omega_L$, is slightly detuned from the $k$-th sideband by $\Delta \omega$:
	 \begin{equation}
		  \omega_L=\omega_{21}-k \nu + \Delta \omega, \label{Eq:mismatch-cond}
	 \end{equation}
	 with $\Delta \omega \ll \nu$.
	Altogether, the Hamiltonian in Eq.~\eqref{Eq:Hinteraction} describes the nonlinear $k$-th sideband coupling $|1,n \rangle \leftrightarrow |2,n-k\rangle$ including a frequency mismatch.

\subsection{Time evolution }
\label{Sec:exactevolution1}

	The time-dependent dynamics of system is governed by the evolution operator $\hat{\mathcal U}(t,t_0)$ which fulfills the standard Schr\"odinger equation
	\begin{align}
		\frac{\partial}{\partial t} \hat{\mathcal{ U}}(t,t_0)=-\frac{i}{\hbar}  \hat H (t)\hat{\mathcal{ U}}(t,t_0)\text{.}
		\label{Eq::dt_U_Operator_DEq_hbar}
	\end{align} 
	Note that in Eq.~\eqref{Eq::dt_U_Operator_DEq_hbar} the factor  $1/\hbar$  always compensates the factor $\hbar$ in the Hamiltonian $\hat H$.
	To ovoid superfluous coefficients we introduce the notation 
	\begin{align}
		\hat H_\text{int}(t)=|\kappa| \hbar \hat {\mathcal{H}}(t)
		\text{,}
	\end{align}
	in terms of the dimensionless Hamiltonian $\hat{\mathcal{H}}$, cf. Eq.~\eqref{Eq:Hinteraction}, which also enables us to track the dependencies on the coupling strength $|\kappa|$ throughout the following.
	
	The formal solution to the reformulated evolution equation
	\begin{align}
		\partial_t\hat{\mathcal{ U}}(t,t_0)=-i |\kappa|  \hat {\mathcal{H}} (t)\hat{\mathcal{ U}}(t,t_0)
		\label{Eq::dt_U_Operator_DEq}
	\end{align}
	can be written in terms of the time ordered exponential 
	\begin{align}
		\hat{\mathcal{ U}}(t,t_0)=\mathcal{T}\exp\left\{-i|\kappa|\int\limits_{t_0}^{t}d\tilde t\, \hat {\mathcal{H}} (\tilde t )\right\}
		\text{,}
		\label{Eq::T_Exp}
	\end{align}
	where the time-ordering prescription $\mathcal{T}$  orders operators with higher $t$ to the right, see for example Refs.~\cite{Schleich,VogelW2006}.
	The time ordered exponential can be represented in terms of
	the Magnus expansion~\cite{Magnus1954,BlanesaCOR2009} 
	\begin{align}	
		&\hat{\mathcal{ U}}( t , t_0)=\exp\left\{-i \hat{\mathcal{M}}( t , t_0) \right\}
		\text{,}
		\label{Eq:MagnusExpansion}
	\end{align}
	where the exponent is the so called Magnus series
	\begin{align}
		-i \hat{\mathcal{M}}( t , t_0)= 
		\sum\limits_{\ell=1}^\infty (-i |\kappa|)^\ell \hat{\mathcal{M}}^{[\ell]}( t , t_0)
		\text{.}
		\label{Eq::Magnus_Series_Operator}
	\end{align}
	
	As in the case of Dyson series~\cite{Dyson1949}, approximations of solutions may be obtained from this representation by truncating the series at any order of the coupling coefficients $|\kappa|$.
	The first order Magnus expansion approximation, given by the first term of the series
	\begin{align}
		\hat{\mathcal{M}}^{[1]}( t , t_0)= \int\limits_{ t_0}^ t  d  t_1\, \mathcal{H}( t_1)
		\text{,}
	\end{align}
	reads as 
	\begin{align}
		\hat{\mathcal{ U}}^{[1]}( t , t_0)=e^{-i|\kappa|\int_{ t_0}^ t  d\tilde  t \, \hat{\mathcal{H}}(\tilde t ) } 
		\label{Eq::U_First_Order_magnus}
	\end{align}
	which corresponds to Eq.~\eqref{Eq::T_Exp} with neglected time ordering, cf.~\cite{KrummV2018}.
	All other terms, $l>1$, of the Magnus series are corrections in terms of time ordering and are therefore referred to as time-ordering effects~\cite{ChristBMS2013,QuesadaS2014,QuesadaS2015,QuesadaS2016,KrummSV2016,KrummV2018,LipfertHPK2018}.
	These time-ordering terms consist of ordered integrals of linear combinations of nested commutators, e.g.,
	\begin{align}
		&\hat{\mathcal{M}}^{[2]}( t , t_0)=\frac{1}{2}\int\limits_{ t_0}^ t  d t_1\int\limits_{ t_0}^{ t_1} d t_2 
		[\hat{\mathcal{H}}( t_2),\hat{\mathcal{H}}( t_1)]\text{,}
	\\
		&\hat{\mathcal{M}}^{[3]}( t , t_0)=\frac{1}{6}\int\limits_{ t_0}^ t  d t_1\int\limits_{ t_0}^{ t_1} d t_2\int\limits_{ t_0}^{ t_2} d t_3
	\notag\\
		&\times\left\{  [\hat{\mathcal{H}}( t_1),[\hat{\mathcal{H}}( t_2),\hat{\mathcal{H}}( t_3)] ]+[\hat{\mathcal{H}}( t_3),[\hat{\mathcal{H}}( t_2),\hat{\mathcal{H}}( t_1)] ]
		\right\}\text{,}
		\label{Eq::Mn_third_order_operator}
	\end{align}
	and they can attain quite complex forms for higher orders, cf.~\cite{PratoL1997}.
	
	For larger time intervals, approximations of increasing orders in terms of the truncated series~\eqref{Eq::Magnus_Series_Operator} do not necessarily converge towards the exact solution of the dynamics~\cite{BlanesaCOR2009}.  
	Thus, it is imperative to take into account such limitations on the convergence when analyzing time-ordering corrections.
	A precise knowledge of convergence bounds usually requires an a priori knowledge of the exact solution. 
	Fortunately, as will be shown below, the Hamiltonian~\eqref{Eq:Hinteraction} belongs to the scarce class of time-dependent physical systems where exact solutions can be derived analytically. 	
	
\section{Explicit solution of the dynamics }
\label{Sec:exactevolution2}

	Bellow we derive a solution to Eq.~\eqref{Eq::dt_U_Operator_DEq} for the Hamiltonian~\eqref{Eq:Hinteraction} that includes all time-ordering effects.
	Based on this we calculate time-dependent density matrices---in Fock-basis representation---of the motional states for different initial configurations. 
	
\subsection{Decoupling and solving the evolution equation by a spinor formalism}
	From the dimensionless Fock basis representation of the Hamiltonian~\eqref{Eq:Hinteraction}  
	\begin{align}
		\hat {\mathcal{H}}( t )= \sum\limits_{n=0}^\infty \omega_n \left[ e^{-i \Delta \omega  t } e^{i \theta}|2,n\rangle \langle 1,n+k|+\text{H.c} \right]
	\notag\\
		\text{with } w_n=
		\cos\left(\Delta\phi+\frac{\pi}{2}k\right)\eta^k e^{-\eta^2/2}\sqrt{\frac{n!}{(n+k)!}}L^{(k)}_n (\eta^2)
		\label{Eq::H_Fock}
	\end{align}
	we can see, that the interaction is entirely
	described in terms of projectors constructed from the states $|2,n\rangle$ and $| 1,n+k\rangle$ with  $n=0,1,\dots$.
	
	A compact notation for such projectors can be formulated in terms of the spinors 
	\begin{align}
		\Psi_n=
		\begin{pmatrix}
			\langle 2,n|e^{-i\theta/2}\\
			\langle 1,n+k|e^{i\theta/2}
		\end{pmatrix}
		\Leftrightarrow
		\Psi_n^\dagger=
		\begin{pmatrix}
			e^{i \theta/2}| 2,n\rangle\\
			e^{-i \theta/2}| 1,n+k\rangle
		\end{pmatrix}^T
		\text{.}
		\label{Eq::Psi_n_Sinor_Def}
	\end{align}
	These spinors fulfill an orthogonality relation 
	\begin{align}
		\Psi_n\Psi_{n'}=\delta_{n,n'}\mathbf{I} \text{, with }\mathbf{I}=
		\begin{pmatrix}
			1	&	0	\\
			0	&	1
		\end{pmatrix}
		\label{Eq::Psi_n_Orthogonality}
	\end{align}
	and allow us to formulate a completeness relation
	as, cf.~\cite{KrummV2018}
	\begin{align}
		\hat{1}	
		&=
		\sum\limits_{n=0}^\infty \Psi_n^\dagger
		\mathbf{I}\Psi_n
		+\sum\limits_{q=0}^{k-1}|1,q\rangle\langle 1,q|		
		\text{.}
	\end{align}
	By applying this completeness relation, the Hamiltonian~\eqref{Eq::H_Fock} can be written in the compact form 
	\begin{align}
		\hat {\mathcal{H}}( t )=\sum\limits_{n=0}^\infty \Psi_n^\dagger
		\mathbf{H}_n( t )
		\Psi_n
		\text{,}
		\label{Eq::H_Spinor}
	\\
		\text{with }\mathbf{H}_n( t )=
		\begin{pmatrix}
			0				&	w_n e^{-i\Delta \omega  t }	\\
			w_n e^{i\Delta \omega t }	&	0
		\end{pmatrix}
		\text{.}
		\label{Eq::H_n}
	\end{align}

	From the quasi-diagonal form of the Hamiltonian in the spinor basis it may be hypothesized that any evolution of the system will also be representable in this basis.
	Thus,  a similarity ansatz  for the evolution operator 
	\begin{align}
		\hat{\mathcal{U}}( t , t_0)=\sum\limits_{n=0}^\infty \Psi_n^\dagger
		\mathbf{U}_n( t , t_0)\Psi_n
		+\sum\limits_{q=0}^{k-1}|1,q\rangle\langle 1,q|
		\label{Eq::U_Spinor}
	\end{align}
	with $\mathbf{U}_n( t , t_0)\in\mathbb{C}^{2{\times}2}$  and initial condition $\left.\mathbf{U}_n( t , t_0)\right|_{ t = t_0}=\mathbf I$ is suitable.
	Substituting~\eqref{Eq::H_Spinor} and~\eqref{Eq::U_Spinor} into~\eqref{Eq::dt_U_Operator_DEq}  decouples the evolution equation 
	in terms of the $2{\times}2$ matrix differential equations 
	\begin{align}
		\partial_ t  
		\mathbf{U}_n( t , t_0)
		=-i |\kappa|
		\mathbf{H}_{n}( t )
		\mathbf{U}_n( t , t_0)		
		\text{.}
		\label{Eq::dt_Un_Matrix_ODE}
	\end{align}
	for $n=0,1, \dots$.
	
	The time-dependent coefficient matrix~\eqref{Eq::H_n} is representable as linear combination of Pauli matrices  that (multiplied by the imaginary unit $i$) generate the Lie-group SU(2). Thus, solutions to~\eqref{Eq::dt_Un_Matrix_ODE} are always representable as 
	\begin{align}
		\mathbf{U}_n( t , t_0)=&\begin{pmatrix}
			a_n( t , t_0)	&	b_n( t , t_0)	\\
			-b^*_n( t , t_0)	&	a^*_n( t , t_0)
		\end{pmatrix}\text{,}
		\label{Eq::Un_with_an_bn}
	\\
		\text{where } & |a_n( t , t_0)|^2+|b_n( t , t_0)|^2=1\text{.}
		\label{Eq::Unitarity_2x2}
	\end{align}
	Note, that in the treatment of parametric down conversion with monochromatic pumps, the solutions to matrix differential equations with parameter dependent coefficients of similar form are known~\cite{Kolobov1999,LipfertHPK2018}.
	
	Essentially, solutions to~\eqref{Eq::dt_Un_Matrix_ODE} can be obtained by transforming the equations into systems with  constant coefficients. Such a system can then be directly solved in terms of matrix exponentials---see Appendix~\ref{Sec::Matrix_Diff_Eq} for a stepwise derivation.
	The explicit solutions then read  
	\begin{align}
		a_n( t , t_0)=e^{-i \Delta\omega [ t - t_0]/2}&
		\bigg[
		\cos(\Gamma_n[ t - t_0])
	\notag\\	
		&+ \frac{i \Delta\omega}{2 \Gamma_n}\sin(\Gamma_n[ t - t_0])
		\bigg]\text{,}
	\notag\\
		b_n( t , t_0)=e^{-i \Delta\omega [ t + t_0]/2}&\frac{|\kappa| w_n}{i \Gamma_n}
		\sin(\Gamma_n[ t - t_0])
		\text{,}
		\label{Eq::an_bn}
	\end{align}
	for  $n=0,1,\dots$
	with $\Gamma_n=\sqrt{\left(\frac{\Delta\omega}{2}\right)^2+w_n^2|\kappa|^2}$.
	In Ref.~\cite{KrummV2018} an approximated solution was found in terms of neglected time ordering---i.e., in terms of~\eqref{Eq::U_First_Order_magnus}. Here, we have found  an analytic expression that incorporates all time-ordering effects, i.e., an explicit representation of the time ordered exponential~\eqref{Eq::T_Exp} for the Hamiltonian~\eqref{Eq:Hinteraction}. 	
	
\begin{figure}[ht]
		\centering
		\includegraphics*[width=8.6cm]{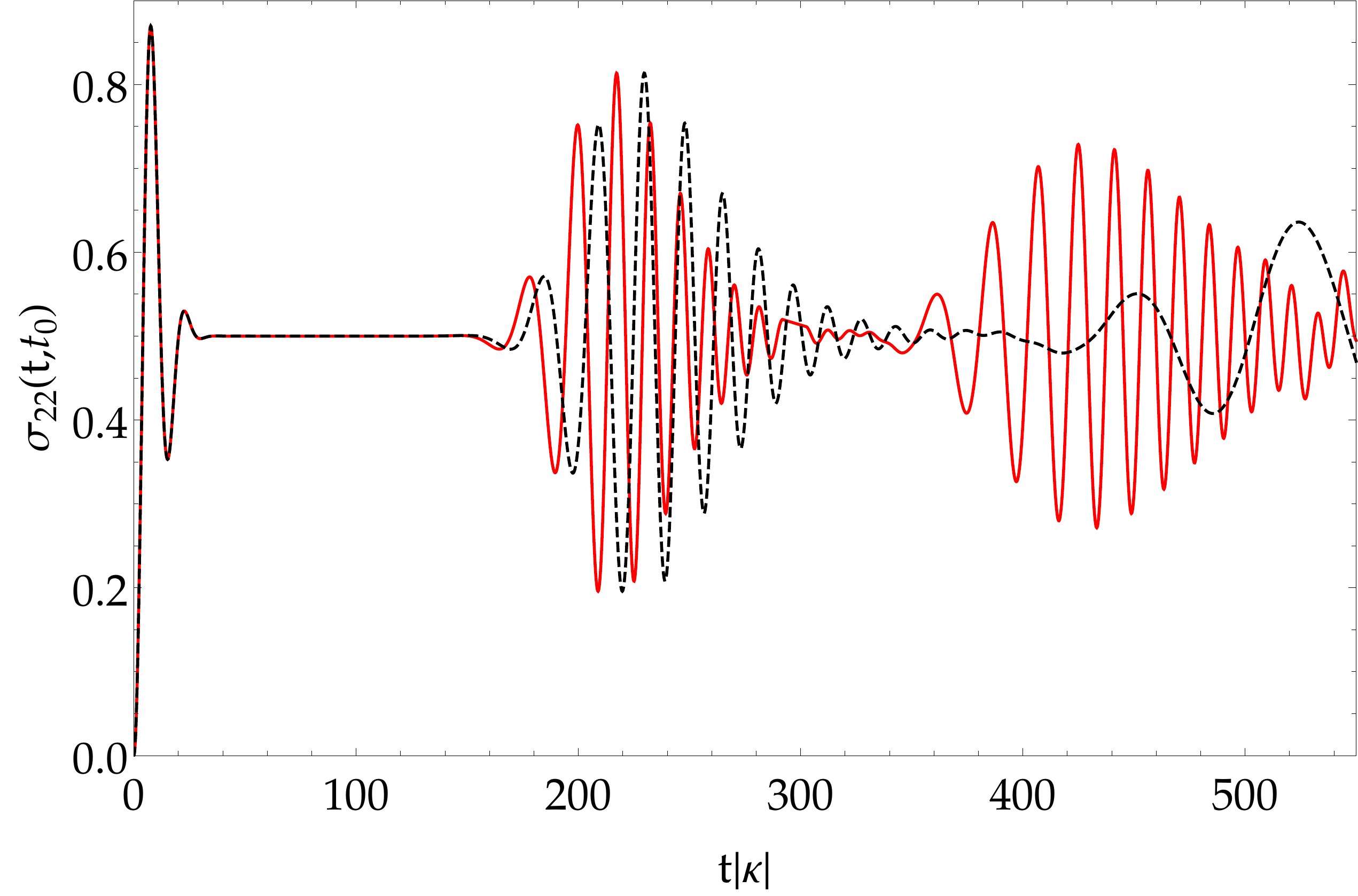}
		\caption{
		The population probability of the excited electronic state $\sigma_{22}(t,t_0)$ as given in Eq.~\eqref{Eq::sigma_22} obtained from the exact analytical solution for $\hat{\mathcal U}$ (red, solid) and for neglected time ordering (black, dashed) with the initial state $\hat\rho^{(1)}(t_0=0)=|1,\alpha_0\rangle\langle1,\alpha_0 |$.
		Here, parameters have been chosen as, $\Delta\omega/|\kappa|=0.005$, $\Delta\phi=0$, $k=2$, and $\eta=0.2$. 	
		This is a verification of the numerical results obtained in Ref.~\cite{KrummV2018}. 
		}\label{Fig::Reproduction_sigma_22}
	\end{figure}

	This analytic expression allows us, e.g., to obtain the dynamics of the population probability of the excited electronic state,
		\begin{align}
			\sigma_{22}(t,t_0)=\sum\limits_{n=0}^\infty\langle 2,n|
			\hat{\mathcal U}(t,t_0)&e^{-\frac{i}{\hbar}\hat H_0(t-t_0)}  \hat\rho(t_0) e^{\frac{i}{\hbar}\hat H_0(t-t_0)}
		\notag\\
			&\times\hat{\mathcal U}^\dagger(t,t_0)| 2,n \rangle
			\label{Eq::sigma_22}
		\end{align}
		and to compare it to the dynamics with neglected time ordering.
		This is illustrated in Fig.~\ref{Fig::Reproduction_sigma_22}.
		Note that this is a reproduction of results that have been obtained numerically in Ref.~\cite{KrummV2018}. 
		Here however all results are based on analytical expressions.
		
\subsection{Density matrices}
		
	Let us apply the derived solution to investigate the evolution of the motional states of the ion.
	For this purpose we calculate the density matrix of the motional state 
	\begin{align}
		\rho_{m,n}(t)=\sum\limits_{j=1,2} \langle j,m |\hat{\mathcal U}(t,t_0)&e^{-\frac{i}{\hbar}\hat H_0(t-t_0)}  \hat\rho(t_0) e^{\frac{i}{\hbar}\hat H_0(t-t_0)}
	\notag\\
		&\times\hat{\mathcal U}^\dagger(t,t_0)| j,n \rangle
	\end{align}
	with the electronic states traced out.
	For the initial coherent state in the electronic ground state, $\hat\rho^{(1)}(t_0)=|1,\alpha_0\rangle\langle1,\alpha_0 |$, and in the excited electronic state $\hat\rho^{(2)}(t_0)=|2,\alpha_0\rangle\langle2,\alpha_0 |$,
	the motional density matrices read as 
	\begin{align}
		\rho_{n,m}^{(1)}(t,t_0)=&
		\frac{\alpha_0^{n+k}\alpha_0^{*(m+k)}e^{-|\alpha_0|^2}}{\sqrt{(n+k)!(m+k)!}} 
		e^{-i\left(\Phi_{1,n+k}-\Phi_{1,m+k}\right)[t-t_0]}
	\notag\\
		\times&
		b_{n}(t,t_0)b^*_{m}(t,t_0)	
	\notag\\
	+&
		\frac{\alpha_0^{n}\alpha_0^{*m}e^{-|\alpha_0|^2}}{\sqrt{n! m!}}	
			e^{-i\left(\Phi_{1,n}-\Phi_{1,m}\right)[t-t_0]}	
	\notag\\
		\times&
			a^*_{n-k}(t,t_0)a_{m-k}(t,t_0)
		\label{Eq::rho_nm_1}
	\end{align}
	and 
	\begin{align}
		\rho^{(2)}_{n,m}(t,t_0)
	&=
		\frac{\alpha_0^{n}\alpha_0^{*m}e^{-|\alpha_0|^2}}{\sqrt{n!m!}}e^{-i\left(\Phi_{1,n}-\Phi_{1,m}\right) [t-t_0]}
	\notag\\
		&\times a_{n}(t,t_0)a_{m}^*(t,t_0)
	\notag\\
		&+\frac{\alpha_0^{n-k}\alpha_0^{*(m-k)}e^{-|\alpha_0|^2}}{\sqrt{(n-k)!(m-k)!}}e^{-i\left(\Phi_{1,n-k}-\Phi_{1,m-k}\right)[t-t_0]}
	\notag\\
		&\times b^*_{n-k}(t,t_0)b_{m-k}(t,t_0)
		\label{Eq:analyticExact-excited}
	\end{align}
	respectively.
	Details regarding the preparation of coherent motional states can be found in Refs.~\cite{M96,Wine90}.
	Here $\Phi_{j,n}=n \nu+[j-1] \omega_{21}$ is the eigenvalue $\hat H_0|j,n\rangle=\hbar\Phi_{j,n}|j,n\rangle$ and we have defined $ a_{m}(t,t_0)=1$ and $ b_{m}(t,t_0)=0$ for negative indices $m<0$.
	
	In the case of neglected time ordering~\eqref{Eq::U_First_Order_magnus} we replace $a_n$ and $b_n$ in~\eqref{Eq:analyticExact-excited} and~\eqref{Eq::rho_nm_1} by the corresponding terms $a_n^{[1]}$ $b_n^{[1]}$ 
	that one obtains in the first order Magnus expansion approximation~\eqref{Eq::U_First_Order_magnus}, i.e.,
	\begin{align}
		a_n^{[1]}(t,t_0)&=\cos\left(w_n |\kappa|(t-t_0) \mathrm{sinc}\left[\frac{(t-t_0)\Delta\omega}{2}\right]\right)
	\notag\\
		b_n^{[1]}(t,t_0)&=-e^{-i\Delta\omega[t+t_0]/2}
	\notag\\
	&\times
		\sin\left(w_n |\kappa|(t-t_0) \mathrm{sinc}\left[\frac{(t-t_0)\Delta\omega}{2}\right]\right)
		\label{Eq:SolutionNoTimeordering}
		\text{.}
	\end{align}
	Note, that this corresponds exactly to the result derived in Ref.~\cite{KrummV2018}.

\section{Quantum time-ordering effects}
\label{Sec:timeOrderingNCL}

	Finally, using the derived results we are able to rigorously investigate the effects of the time-ordering contributions, $\hat{\mathcal{M}}^{[n>1]}( t , t_0)$ in Eq.~\eqref{Eq:MagnusExpansion}, on the nonclassicality of the system.
	The phrase \textit{nonclassicality} is defined as follows:
	Noting that the density operator of a system (or of a subsystem) can be expressed as a mixture of coherent states using the Glauber-Sudarshan $P$-representation~\cite{Glauber1963,Sudarshan1963},
	\begin{align}
		 \hat \rho(t)= \int d^2\alpha P(\alpha;t) |\alpha \rangle \langle \alpha |,
	\end{align}
	the system is classical if the density operator can be expressed as a classical mixture of coherent states.
	
	The latter means that $P(\alpha;t) \geq 0$.
	That is, the coherent states serve as reference states to divide classicality from nonclassicality.
	If the $P$~functions attains negative values (in the sense of distributions) the state is referred to as nonclassical as it cannot be expressed as a classical mixture of coherent states and hence, 
	it consists of superpositions of them, which is a clear signature of nonclassical behavior~\cite{TG65,Mandel86}.
	
	However, the $P$~function is highly singular for many states and hence, it cannot be observed in experiments.
	To circumvent this issue a regularization procedure was established~\cite{Kiesel10} which converts the $P$~function into a well-behaved quasiprobability $P_\Omega$.
	Due to an appropriate choice of filter functions the latter does only attain negative values if the $P$~function does.
	That is, if the regularized version reveals negativities, then the state is referred to as nonclassical state.
	The applicability was verified in several experiments, see for example Refs.~\cite{Kiesel11,Agudelo15}.
	
	As presented in Ref.~\cite{KrummV2018}, the regularized $P$~function can be calculated out of the, possibly reduced, density matrix elements in Fock basis via
	\begin{align}
		P_\Omega(\alpha;t)= \sum_{m,n=0}^\infty \rho_{m,n}(t) P_{\Omega,n,m}(\alpha),
	\end{align} 
	where  $P_{\Omega,n,m}(\alpha)$ needs to be calculated only once and can be subsequently combined with the density matrix elements $\rho_{m,n}(t)= \langle m | \hat \rho(t) | n \rangle$ for arbitrary times.
	The elements of the regularized $P$~function are calculated via
	\begin{align}
		 \label{Eq:PregulFock}
		&P_{\Omega,n,m}(\alpha) = \frac{16}{\pi^2}w^2  e^{i (n-m)\varphi_\alpha}  \int_0^1 dz  \Lambda_{n,m}(2wz)  z \nonumber \\
		&\times  J_{n-m}(4 w |\alpha|z) \left[ \arccos(z)-z\sqrt{1-z^2} \right],
	\end{align}
	with $J_n(x)$ being the Bessel functions of the first kind, $\alpha = |\alpha| e^{i \varphi_\alpha}$, and 
	\begin{align}
	 \Lambda_{n,m}(x)=
	 \begin{cases}
	    (-x)^{m-n} \sqrt{\frac{n!}{m!}} L_n^{(m-n)}(x^2) &  m \geq n \\
	    x^{n-m} \sqrt{\frac{m!}{n!}} L_m^{(n-m)}(x^2) &  m < n .
	 \end{cases}
	\end{align}
	In Eq.~\eqref{Eq:PregulFock} a radial symmetric filter was assumed.
	Details on appropriate filter functions can be found in Ref.~\cite{Kuehn14}.
	
	\begin{figure*}[ht]
	\centering
	\includegraphics*[width=8.6cm]{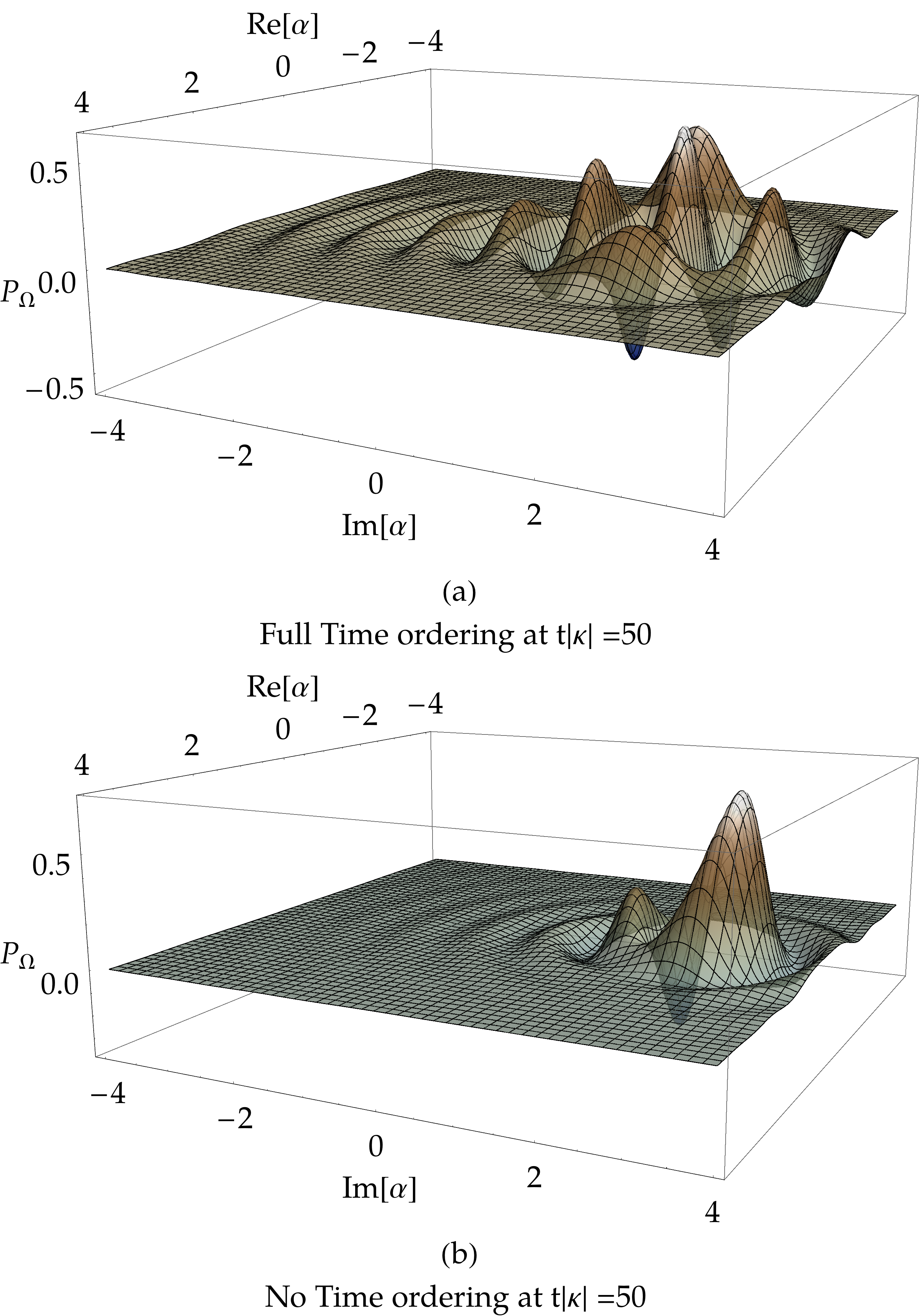}
	\includegraphics*[width=8.6cm]{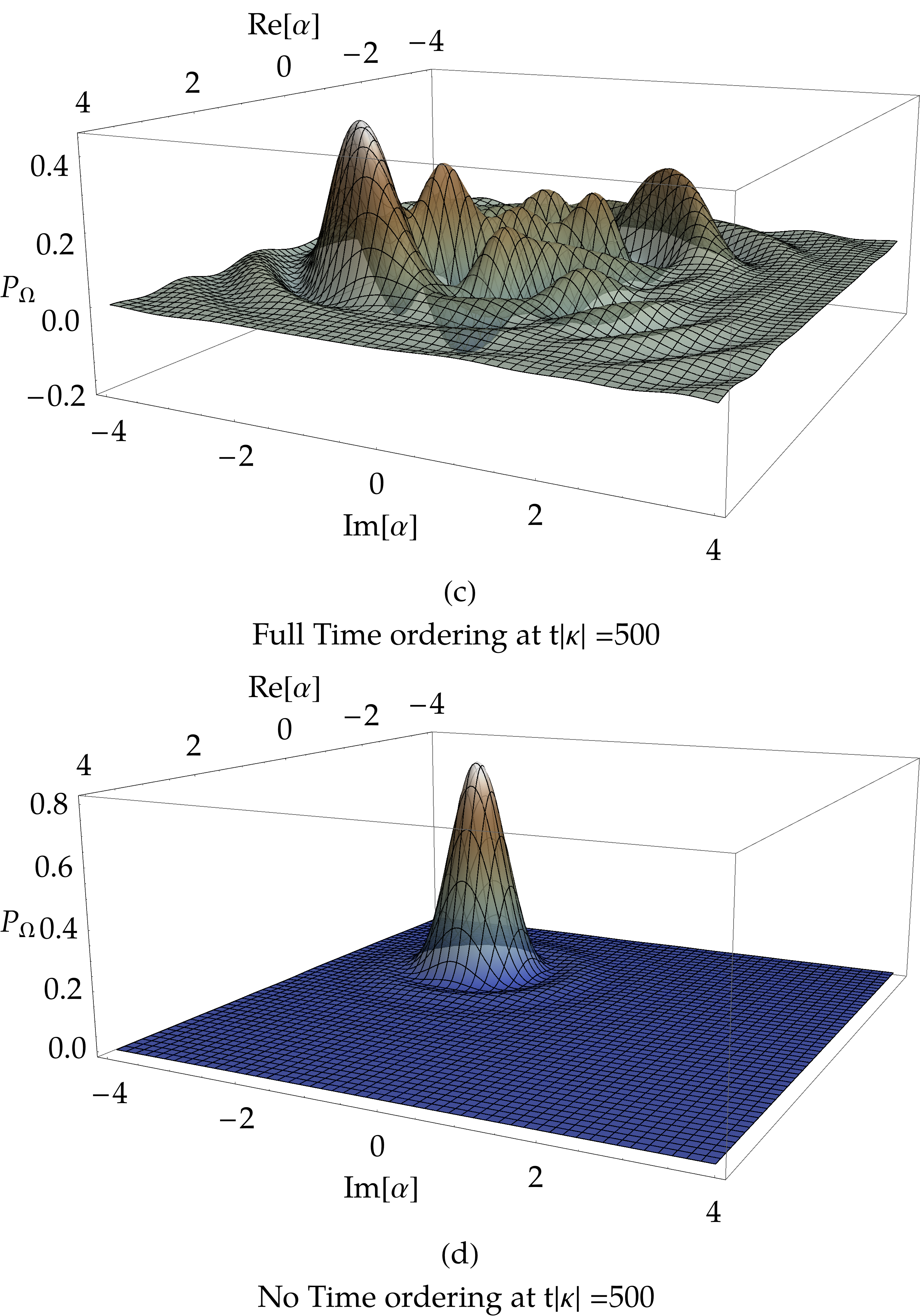}
	\caption{
		The regularized Glauber-Sudarshan $P$~function for $t|\kappa|=50$ in (a) and (b), and $t|\kappa|=500$ in (c) and (d), for the excitation to the second sideband, $k=2$ and $\Delta \Phi=0$ in Eq.~\eqref{Eq:operatorfuncf}.
		Here an electronic excited input state was chosen.
		The two figures (a) and (c) correspond to the analytic, exact solution in Eq.~\eqref{Eq:analyticExact-excited}.
		Figures (b) and (d) are obtained via the solutions without time-ordering effects [Eq.~\eqref{Eq:SolutionNoTimeordering}].
		Parameters: $\eta=0.2,\ \alpha_0=\sqrt{5},$ and $\Delta\omega/|\kappa|=0.1$.
	}\label{Fig:NCL}
	\end{figure*}
	The plots of the quasiprobability in Eq.~\eqref{Eq:PregulFock} are depicted in Fig.~\ref{Fig:NCL} for $t |\kappa| = 50$ and $t |\kappa|=500$,
	leading to differences in figures (a), (b) and (c), (d), respectively, due to different Magnus orders.
	We see that they have a crucial impact on the dynamics of the system.
	
	Keeping in mind that the regularized $P$~function visualizes the state under study in the phase space, one can see that
	Fig.~\ref{Fig:NCL} (a) gives rise to a superposition of two coherent states of unequal amplitude, where the strong negativities indicate quantum interferences.
	However, Fig.~\ref{Fig:NCL} (b) does not show this behavior.
	The differences become even stronger if one considers larger times, see figures (c) and (d).
	In the latter we can hardly verify nonclassicality.
	Thus, on larger time scales the whole nonclassical character is contained in the time-ordering effects.
	
	Note that similar results are obtained using other parameters. 
	On short time scales the time-ordering effects do not have a decisive impact on nonclassicality, see for example~\cite{KrummSV2016}.
	However, on larger time scales, the nonclassical character of the states is governed by non-commuting Hamiltonians.

\section{Time-ordering corrections }
\label{Sec:TOcorrections}

	One may now be interested in how different orders of time ordering affect the  temporal evolution of the system under study.
	Hence, based on the explicit solution, a generating function for arbitrary order terms of the Magnus series is defined.
	 This function allows us to perform an extensive study of the convergence of time-ordering corrections.
	
\subsection{A generating function for time-ordering corrections}	
	Using the orthogonality~\eqref{Eq::Psi_n_Orthogonality} of our spinor formalism~\eqref{Eq::Psi_n_Sinor_Def}
	it is easy to show that the non-equal time commutators of~\eqref{Eq::H_Spinor} fulfill 
	\begin{align}
		[\hat{\mathcal H}( t ),\hat{\mathcal H}( t' )]=
		\sum\limits_{n=0}^\infty 
		\Psi_n^\dagger
		[\mathbf{H}_n( t ),\mathbf{H}_n( t' )]
		\Psi_n
		\text{.}
	\end{align}
	Consequently, there is a one-to one correspondence to the non-equal time commutators of the matrices $\mathbf{H}_n$.
	Thus it follows, that the $\ell$-th order Magnus expansion approximation to the solution of~\eqref{Eq::dt_U_Operator_DEq} corresponds exactly to the result one obtains  
	by evaluating the $\ell$-th  order Magnus expansion approximation to the solution of Eq.~\eqref{Eq::dt_Un_Matrix_ODE}.
	
	The analytic solutions in Eq.~\eqref{Eq::an_bn} allow for the explicit formulation of the full Magnus series.
	This means, we can bring~\eqref{Eq::Un_with_an_bn} in exponential form 
	as
	\begin{align}
		\mathbf{U}_n(|\kappa|)=\exp\left\{-i{\mathbf M}_n(|\kappa|)\right\}
		\text{.}
	\end{align}
	Here and in the remainder of this section we drop the time-dependence of the matrices from our notation and consider the appearing parameters as functions of the coupling parameter $|\kappa|$.
	
	Evaluating the matrix exponential under application of the unitary condition~\eqref{Eq::Unitarity_2x2} shows that 
	the matrix-exponent can be chosen in the form
	\begin{align}
		{\mathbf M}_n(|\kappa|)=&
		\frac{\,\mathrm{arccos}(\mathrm{Re}(a_n(|\kappa|)))}{\sqrt{1-\mathrm{Re}^2(a_n(|\kappa|))}}
	\notag\\
		&\times\begin{pmatrix}
			-\mathrm{Im}(a_n(|\kappa|))	&	i b_n(|\kappa|)	\\
			-i b^*_n(|\kappa|)			&	\mathrm{Im}(a_n(|\kappa|))
		\end{pmatrix}
		\text{.}
		\label{Eq::Mn_Matrix_Exponents}
	\end{align}
	This matrix can now serve as a generating function for the different orders of time-ordering corrections, i.e.,
	a Taylor series expansion of $\mathbf{M}_n(|\kappa|)$ in terms of $|\kappa|$ around $|\kappa|=0$ yields the Magnus series as
	\begin{align}
		 \mathbf{M}_n(|\kappa|)=i\sum\limits_{\ell=1}^\infty(-i|\kappa|)^\ell \mathbf{M}_n^{[\ell]}
		\text{,}
		\label{Eq::Mn_Magnus_Series_Matrix}
	\end{align}
	where 
	\begin{align}
		\mathbf{M}_n^{[\ell]}=\frac{i^{\ell-1}}{\ell!}\left.\frac{d^\ell \mathbf{M}_n(|\kappa|)}{d|\kappa|^\ell}\right|_{|\kappa|=0} 
		\text{.}
	\end{align}
	We have verified
	the equivalence of these expressions with those obtained from nested commutators, cf.~\eqref{Eq::Mn_third_order_operator},  up to 5th order---see Appendix~\ref{App::MagnusTerms}---using expressions in References~\cite{BlanesaCOR2009} and~\cite{PratoL1997}.
	This allows us to obtain time-ordering corrections to arbitrary order.
	
	We want to point out that the Magnus series~\eqref{Eq::Mn_Magnus_Series_Matrix} may not always converge~\cite{BlanesaCOR2009}.
	However, one can show that the Magnus series~\eqref{Eq::Mn_Magnus_Series_Matrix}
	converges for~\cite{BlanesaCOR2009} 
	\begin{align}
		|\kappa|\int\limits_{t_0}^t d\tilde t\, || 
		\mathbf{H}_{n}( \tilde t )||_2=|\kappa w_n|[t-t_0]<\pi
		\label{Eq::Upper_Bound_Time}
	\end{align}
	where $||\cdots||_2$ denotes the spectral norm~\cite{HornJ1985}. 
	In terms of the full operator~\eqref{Eq::U_Spinor} this means, we can guarantee convergence, as long as 
	$w_\text{max}|\kappa|[t-t_0]<\pi$  with $w_\text{max}=\max\limits_{n=0,1,\dots}|w_n|$, also cf. Eq.~\eqref{Eq::H_Fock}.
	
\begin{figure}[ht]
		\includegraphics[width=8.6cm]{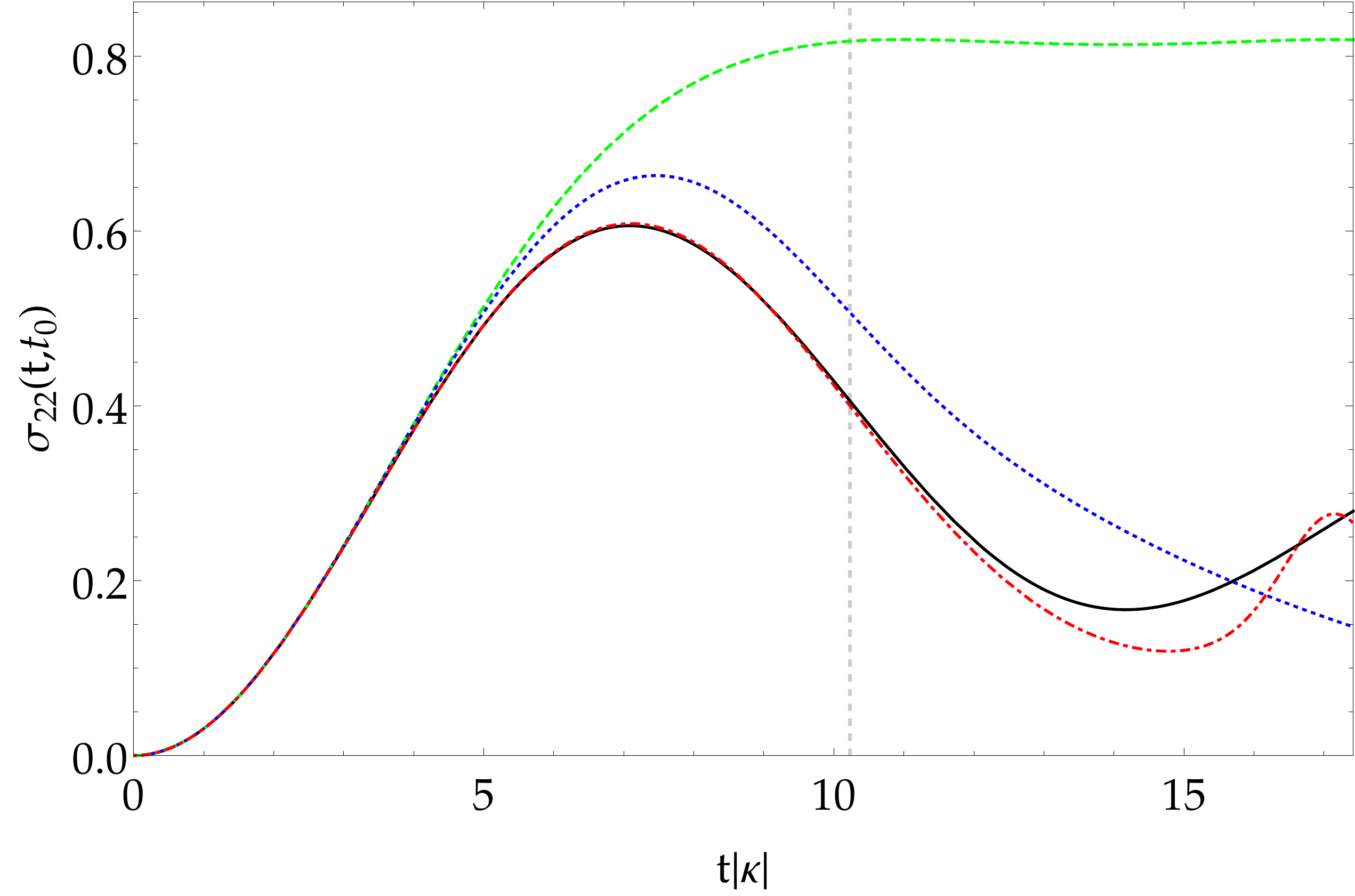}
		\caption{
			The population probability of the excited electronic state $\sigma_{22}(t,t_0)$ as given in Eq.~\eqref{Eq::sigma_22} obtained from the exact solution for $\hat{\mathcal U}$ (black, solid), from the first order approximation , i.e., neglected time ordering (green, dashed), from the second order approximation (blue, dotted), and from the fifth order approximation (red, dot-dashed) with the initial state $\hat\rho^{(2)}(t_0=0)=|2,\alpha_0\rangle\langle2,\alpha_0 |$ in the range $0\leq t|\kappa| <17.4$.
			Here, parameters have been chosen as, $\Delta\omega/|\kappa|=0.224$, $\Delta\phi=\pi/4$, $k=3$, and $\eta=0.4$. 
			This implies, that $w_\text{max}=0.307$.
			A vertical gridline (grey dashed) at $t|\kappa|=\pi/w_{\text{max}}$ marks the region in which the sufficient condition [Eq.~\eqref{Eq::Upper_Bound_Time}] guarantees the convergence of the time-ordering corrections.
			In Section~\ref{Sec::convergence_analysis} we develop criteria that allow us to state that the time-ordering corrections converge in the full displayed range.  
		}\label{Fig::sigma_22_Magnus}
	\end{figure}

	The upper bound~\eqref{Eq::Upper_Bound_Time} is nonetheless only a sufficient criterion for convergence.  
	In Fig.~\ref{Fig::sigma_22_Magnus} we give an illustrative example of convergence well above this upper bound. Here we compare the population probability of the excited electronic state~\eqref{Eq::sigma_22} in different orders of time-ordering corrections to the exact solution.
	The quality of approximation seemingly also improves above the bound~\eqref{Eq::Upper_Bound_Time}.

\subsection{Convergence analysis}\label{Sec::convergence_analysis}
	
	We have seen that the sufficient convergence criterion~\eqref{Eq::Upper_Bound_Time} may underestimate the actual limits of convergence of Magnus expansion approximations.
	As we treat here a frequency-mismatch system one may wonder 
	why the frequency-mismatch $\Delta\omega$ does not appear in the criterion~\eqref{Eq::Upper_Bound_Time}. 
	Sharper bounds seem desirable in the treatment of time-ordering corrections and knowing the analytic expression for the exponents~\eqref{Eq::Mn_Matrix_Exponents} indeed allows 
	us to perform a more sophisticated analysis.
	
	For these purposes we may substitute $w_n\mapsto\tau_n/(|\kappa|[t-t_0])$ and $\Delta\omega\mapsto 2\Lambda/[t-t_0]$ with the dimensionless parameters $\tau_n$ and $\Lambda$ into~\eqref{Eq::an_bn}.
	In this manner we  find expressions of the form
	\begin{align}
		a_n(|\gamma|)& \mapsto\tilde a_\Lambda(\tau_n)	& b_n(|\gamma|)&\mapsto \tilde b_\Lambda(\tau_n)
	\notag\\
		\mathbf{M}_n(|\gamma|)&\mapsto \mathbf{\tilde M}_\Lambda (\tau_n)
		\text{.}
		\label{Eq::aLambda_bLambda_MLambda}
	\end{align}
	Applying the chain rule, it is now easy to show that
	\begin{align}
		\mathbf{M}^{[\ell]}_n=\frac{\tau_n^\ell}{|\kappa|^\ell}\mathbf{\tilde M}_\Lambda^{[\ell]}
		\text{,}
	\end{align}
	with the partial derivatives at $\tau_n=0$
	\begin{align}
		\mathbf{\tilde M}_\Lambda^{[\ell]}=
		\frac{i^{\ell-1}}{\ell!}\left.\frac{\partial^\ell \mathbf{\tilde M}_\Lambda(\tau_n)}{\partial\tau_n^\ell}\right|_{\tau_n=0} .
	\end{align}
	
	Thus it follows, that the 
	Magnus series~\eqref{Eq::Mn_Magnus_Series_Matrix}
	only converges if the
	Maclaurin series
	\begin{align}
		 \mathbf{\tilde M }_\Lambda(\tau_n)=i\sum\limits_{\ell=1}^\infty(-i\tau_n)^\ell \mathbf{\tilde M}_\Lambda^{[\ell]}
		 \label{Eq::M_Lambda_Series}
	\end{align}
	converges.
	To analyze the convergence of the series~\eqref{Eq::M_Lambda_Series}, we consider the matrix elements of $\mathbf{\tilde M}_{\Lambda}(\tau_n)$ as complex functions by replacing $\tau_n\mapsto z$. 
	It is then possible to determine the radii $r_\Lambda$ of convergence~\cite{Needham1997} of these series, $|z|<r_\Lambda$, in terms of the singularities of the analytical expressions $\mathbf{\tilde M}_\lambda(z)$ in the complex plane~\cite{BlanesaCOR2009}. 
	
	In this manner we can obtain exact limits of convergence for the Magnus series.
	Details on this rather elaborate procedure are given in Appendix~\ref{App::Convergence_Analysis}.
	The result reads, see Eq.~\eqref{Eq:resultAppB},
	\begin{align}
		|\tau_n|=|\kappa w_n|[t-t_0]<r_\Lambda=r_{\Delta \omega[t-t_0]/2}
		\text{.}
		\label{Eq::Upper_Limit_exact}
	\end{align}
	
\begin{figure}[ht]
		\includegraphics[width=8.6cm]{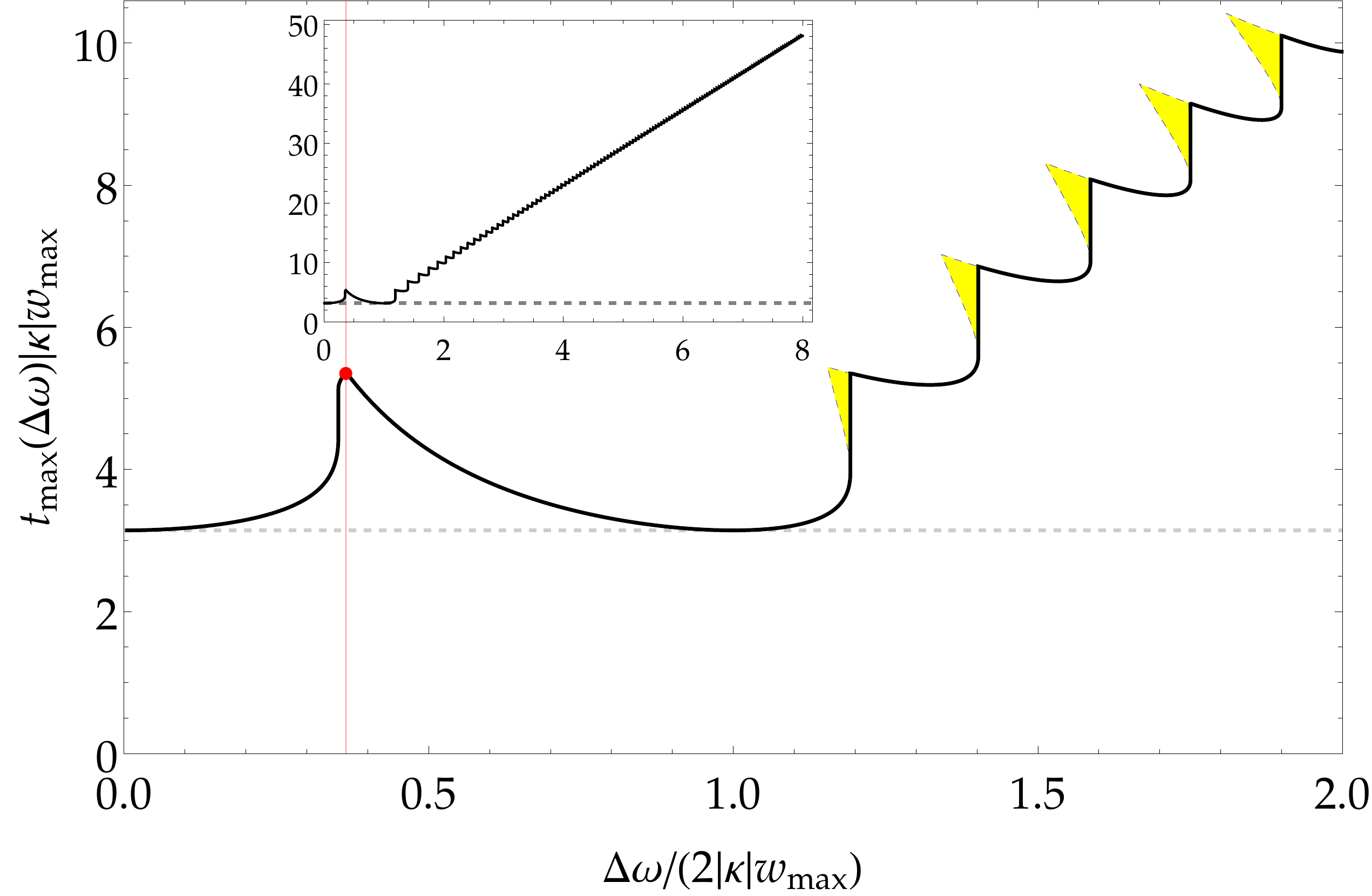}
		\caption{
		The upper bound (black)  $t_\text{max}(\Delta\omega)$  of the convergence time of the Magnus expansion as a function of the frequency-mismatch $\Delta\omega$ as defined in~\eqref{Eq::Upper_Bound_exact}---an inset illustrates its global behavior. 
		The constant upper bound estimate (gray, dashed) as defined in~\eqref{Eq::Upper_Bound_Time}. 
		In the displayed regions (yellow, blue dashed boundary) the Magnus expansion also converges but divergences have appeared at earlier times, i.e., at the upper bound. 
		A vertical gridline (red) and a point (red) mark the position of the first local maximum of $t_\text{max}(\Delta\omega)$. 
		}
		\label{Fig::Upper_Bound}
	\end{figure}

	Note that again $w_n$ appears only as a factor here, thus for the full Magnus expansion of the full operator~\eqref{Eq::U_Spinor} we may define $w_\text{max}=\max\limits_{n=0,1,\dots} |w_n|$.
	Then the Magnus expansion for~\eqref{Eq::U_Spinor} converges for 
	\begin{align}
		0\leq [t-t_0]<t_\text{max}(\Delta\omega) =\min_{\tilde t\in \mathbb{R}_+:|\kappa w_\text{max}\tilde t|=r_{\Delta\omega \tilde t /2}} \tilde t
		\text{,}
		\label{Eq::Upper_Bound_exact}
	\end{align}
	where $t_\text{max}(\Delta\omega)$ is a function of the frequency-mismatch.
	This exact upper bound of convergence is displayed in Fig.~\ref{Fig::Upper_Bound}.
	
	Note that it may happen that the series becomes again convergent for $t>t_\text{max}(\Delta\omega)$ in regions of $t$ where Eq.~\eqref{Eq::Upper_Limit_exact} is fulfilled. 
	However, $t_\text{max}(\Delta\omega)$ is the exact upper limit of continuous convergence.  Judging from the display of these regions in Fig.~\ref{Fig::Upper_Bound}, they do not increase the range of convergence significantly.
	For a large frequency-mismatch the convergence time increases in a linear fashion.  Note that too large detunings may undermine the validity of the model in Eq.~\eqref{Eq:Hinteraction}.
	
	Based on this analysis, the frequency-mismatch $\Delta\omega$  in Fig.~\ref{Fig::sigma_22_Magnus}  has been chosen such that
	the maximal displayed value $t$  corresponds to the first local maximum of $t_\text{max}(\Delta\omega)$ in Fig.~\ref{Fig::Upper_Bound}. 
	Thus, we can guarantee convergence of the time-ordering correction to the exact solution for all ranges of $t$ displayed in Fig.~\ref{Fig::sigma_22_Magnus}.

\section{Non-classical pump fields }
\label{Sec:NCLpump}

	As thoroughly discussed in this work, the model studied was solved in Ref.~\cite{KrummV2018} via an alternative approach.
	In this Section we reconsider each solution and add some remarks regarding the strengths of each strategy. 
	Note that even if in this contribution and in Ref.~\cite{KrummV2018} the same physical model is considered, the investigated scenarios clearly differ.
	
	In Ref.~\cite{KrummV2018} the Hamiltonian, which is  given in Eq.~\eqref{Eq:Hinteraction}, was solved via  quantization of the pump field.
	Thus, the Hilbert space was extended and the Hamiltonian became time-independent in the Schr\"odinger picture.
	This procedure corresponds formally to the replacement
	\begin{align}
		\kappa e^{-i \omega_L t} \rightarrow \hat b  |\kappa'|.
	\end{align}
	The time-ordering effects, discussed in this work, were naturally contained in the straightforward solution of the time-independent Hamiltonian. 
	The extension of the Hilbert space yields  more cumbersome algebraic expressions for all observables under investigation.
	Additionally, the convergence to the semi-classical solution was only presented via numerical solutions obtained by the \textsc{python} package \textsc{qutip}~\cite{Qutip1,Qutip2}.
	However, the approach has the advantage of more general input pump fields.
	As the pump is treated in a quantized manner in its own Hilbert space, one can consider the scenario for arbitrary input states of the pump.
	In the case of a strong coherent input field the semi-classical solution is obtained on a finite time scale.
	Nevertheless, the consideration of squeezed or cat-like input states is possible without significant additional effort.
	
	As an example one may assume the following input state for the pump field:
	\begin{align}
	 \label{Eq.NCLpumpt1}
	 \hat \rho_\text{pump}= |\Psi_\text{SV} \rangle \langle \Psi_\text{SV} |
	\end{align}
	with the squeezed vacuum state
	\begin{align}
	\label{Eq.NCLpumpt2}
	 |\Psi_\text{SV} \rangle = \frac{1}{\cosh \xi} \sum_{n=0}^\infty (- \tanh \xi)^n \frac{\sqrt{(2n)!}}{2^n n!} | 2n \rangle.
	 \end{align}
	The required algebra is explicitly given in  Ref.~\cite{KrummV2018}.
	On this basis one may calculate the reduced density matrix of the motional state and  the resulting regularized $P$~function, cf. Eq.~\eqref{Eq:PregulFock}.
	\begin{figure}[h]
		\centering
		\includegraphics*[width=8.6cm]{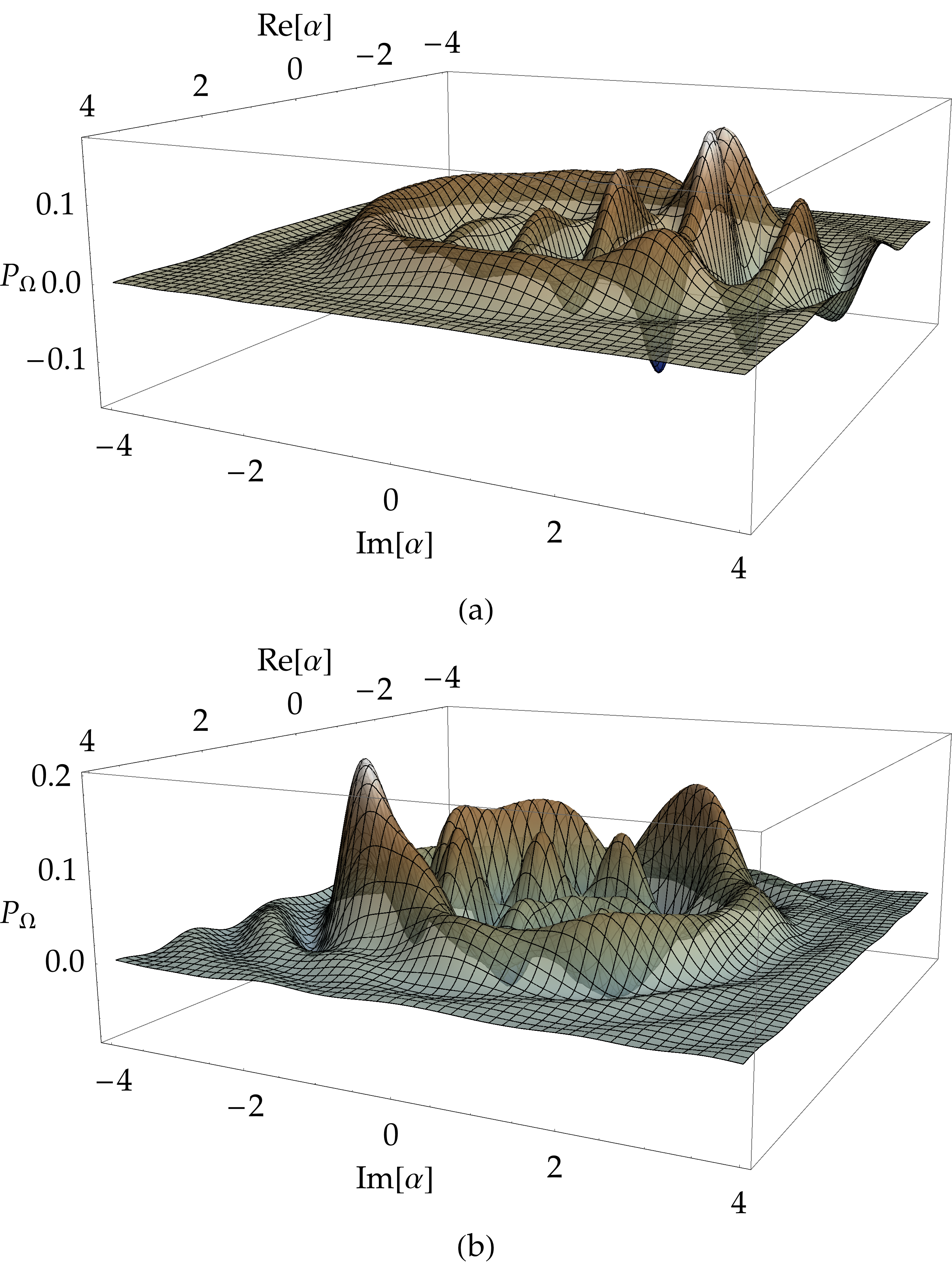}
		\caption{
		The regularized Glauber-Sudarshan $P$~function for $t|\kappa'|=50$ (a) and $t|\kappa'|=500$ (b) for the same parameters as in Fig.~\ref{Fig:NCL} 
		but for a nonclassical input pump field as given in Eqs.~\eqref{Eq.NCLpumpt1} and~\eqref{Eq.NCLpumpt2}, for $\xi=2$.
		}\label{Fig:NCL-pump}
	\end{figure}
	The results are depicted in Fig.~\ref{Fig:NCL-pump}, where we plotted the regularized $P$~function for $t |\kappa'|=50$ and $t |\kappa'|=500$, considering a squeezing parameter of $\xi=2$.
	This choice of $\xi$ corresponds to a squeezing strength of approximately $17.4$ dB, so far $15$ dB squeezing is experimentally available~\cite{squeez}.
	We use $\xi=2$ to visualize more significant effects of a nonclassical pump field.
	The presented quasiprobabilities in Fig.~\ref{Fig:NCL-pump} are similar to those for the classical input, cf. Fig.~\ref{Fig:NCL} (a) and (c).
	The regularized $P$~functions in Fig.~\ref{Fig:NCL-pump}, however, exhibit an additional circular distribution surrounding the structures as they typically occur for a the semi-classical pump.
	Hence, the usage of a nonclassical pump might  lead to so far unstudied effects which need a more detailed treatment, which is beyond the scope of this work.

\section{Summary and Conclusion }
\label{Sec:conclusion}

	In summary, we derived an exact solution for the dynamics corresponding to the classically driven detuned nonlinear Jaynes-Cummings model describing the quantized motion of a trapped ion.
	The solution was formulated by applying a spinor formalism which decoupled the dynamics in terms of $2\times2$ matrix differential equations.
	These matrix differential equations have been solved analytically, which resulted in analytical expressions for the evolution operator of  the system.
	Applying the latter we have reproduced results on the dynamics of the  population probability of the excited electronic state, that have be obtained numerically in Ref.~\cite{KrummV2018}.
	
	The analytical expression for the time-evolution operator allows for the investigation of time-ordering effects. 
	Using the regularized version of the Glauber-Sudarshan $P$~function, we have discussed the influence of the latter on the nonclassicality of the motional states of the ion.
	Especially on large time-scales the whole nonclassical character is contained in contributions which are connected to non-commuting Hamiltonians.
	
	Furthermore, based on the  analytical solution, a generating function for the time-ordering corrections could be derived. 
	This function generates all terms up to arbitrary order of the Magnus expansion for the evolution operator of the  nonlinear Jaynes-Cummings model.
	As an example for the impact of different orders of time-ordering corrections on observables the population probability of the excited electronic state has been considered without time ordering, with full time ordering, and with different orders of time-ordering corrections.
	With increasing orders of corrections the quality of approximation improved.
	By extending the generating function to the complex plane, we could determine the convergence of time-ordering corrections by locating the singularities of the former in the complex plane.
	It was shown, that these exact upper bounds depend on the frequency-mismatch and exceed known sufficient upper bounds over a wide range of detunings. 
	Additionally we have observed isolated regions of convergence above these upper bounds.
	The obtained exact upper bound has allowed us to analyze the impact of time-ordering corrections above the sufficient upper bound.
	
	In the last part of the work we discussed the influence of a nonclassical pump field on the motional state.
	We used a squeezed vacuum state to reveal discrepancies to the semi-classical solutions in phase-space. 
	The findings suggest that a nonclassical pump causes a dynamics beyond the one derived in the semi-classical-pump scenario.

\section*{Acknowledgments}
	T.L., M.I.K. and W.V. acknowledge funding from the European Union's
	Horizon 2020 Research and Innovation program under Grant
	Agreement No. 665148 (QCUMbER).
\appendix
\section{Solution to the matrix differential equation}\label{Sec::Matrix_Diff_Eq}
	To solve the matrix differential equations 
	\begin{align}
		\partial_ t  
		\mathbf{U}_n( t , t_0)
		=-i  
		\begin{pmatrix}
			0				&	|\kappa| w_n e^{-i\Delta\omega  t }	\\
			|\kappa| w_n e^{i\Delta\omega  t }	&	0
		\end{pmatrix}
		\mathbf{U}_n( t , t_0)
	\end{align}
	with initial condition  $\left.\mathbf{U}_n( t , t_0)\right|_{ t = t_0}=\mathbf{I}$ 
	we write
	\begin{align}
		\begin{pmatrix}
			0				&	|\kappa| w_n e^{-i\Delta\omega  t }	\\
			|\kappa| w_n e^{i\Delta\omega  t }	&	0
		\end{pmatrix}
		=
		\mathbf{S}^\dagger( t )
		\begin{pmatrix}
			0				&	|\kappa| w_n 	\\
			|\kappa| w_n 	&	0
		\end{pmatrix}
		\mathbf{S}( t )
		\text{,}
	\notag\\
	\text{with }
		\mathbf{S}( t )=
		\begin{pmatrix}
			e^{i\Delta\omega  t  /2}	&	0	\\
			0		&	e^{-i\Delta\omega t /2}
		\end{pmatrix}
	\end{align}
	such that 
	\begin{align}
		\mathbf{S}( t )
		\partial_ t  
		\mathbf{U}_n( t , t_0)
		=-i  
		\begin{pmatrix}
			0				&	|\kappa| w_n 	\\
			|\kappa| w_n 	&	0
		\end{pmatrix}
		\mathbf{S}( t )
		\mathbf{U}_n( t , t_0)
		\text{}
	\end{align}
	Adding the term $\left[\partial_ t \mathbf{S}( t )\right]
		\mathbf{U}_n( t , t_0)$ on both sides of the equation, applying the product rule on the left hand side, 
		and executing the derivative 
	\begin{align}
		\partial_ t \mathbf{S}( t )=-i
		\begin{pmatrix}
		    -\Delta\omega/2	&	0	\\
		    0		&	\Delta\omega/2
		\end{pmatrix}
		\mathbf{S}( t )
	\end{align}
	on the right hand side yields the constant parameter differential equation
	\begin{align}
		\partial_ t 
		\left[
		\mathbf{S}( t )
		\mathbf{U}_n( t , t_0)\right]
		=-i  
		\begin{pmatrix}
			-\Delta\omega/2				&	|\kappa| w_n 	\\
			|\kappa| w_n 	&	\Delta\omega/2
		\end{pmatrix}
		\left[\mathbf{S}( t )
		\mathbf{U}_n( t , t_0)
		\right]
		\text{,}
	\end{align}
	with initial condition $\left.\left[\mathbf{S}( t )
	\mathbf{U}_n( t , t_0)\right]\right|_{ t = t_0}=\mathbf{S}( t_0)$.
	The solution is easily found as 
	\begin{align}
		\left[\mathbf{S}( t )
		\mathbf{U}_n( t , t_0)
		\right]=\exp\left\{-i( t - t_0)\begin{pmatrix}
			-\Delta\omega/2				&	|\kappa| w_n 	\\
			|\kappa| w_n 	&	\Delta\omega/2
		\end{pmatrix}\right\}
		\mathbf{S}( t_0)
	\end{align}
	which leads to
	\begin{align}
		\mathbf{U}_n( t , t_0)
		&=\mathbf{S}^\dagger( t )\exp\left\{-i( t - t_0)\begin{pmatrix}
			-\Delta\omega/2				&	|\kappa| w_n 	\\
			|\kappa| w_n 	&	\Delta\omega/2
		\end{pmatrix}\right\}
		\mathbf{S}( t_0)
	\notag\\
		&=
		\begin{pmatrix}
			a_n( t , t_0)	&	b_n( t , t_0)	\\
			-b^*_n( t , t_0)	&	a^*_n( t , t_0)
		\end{pmatrix}
		\text{,}
	\end{align}
	with
	\begin{align}
		a_n( t , t_0)=e^{-i \Delta\omega [ t - t_0]/2}&
		\bigg[
		\cos(\Gamma_n[ t - t_0])
	\notag\\	
		&+ \frac{i \Delta\omega}{2 \Gamma_n}\sin(\Gamma_n[ t - t_0])
		\bigg]
	\notag\\
		b_n( t , t_0)=e^{-i \Delta\omega [ t + t_0]/2}&\frac{|\kappa| w_n}{i \Gamma_n}
		\sin(\Gamma_n[ t - t_0])
		\text{,}
	\end{align}
	and $\Gamma_n=\sqrt{\left(\frac{\Delta\omega}{2}\right)^2+w_n^2|\kappa|^2}$.

\section{Magnus Terms}\label{App::MagnusTerms}
	The Magnus terms up to fifth order have been computed from the generating function~\eqref{Eq::Mn_Matrix_Exponents} and the ordered nested non-equal-time commutators, cf.\eqref{Eq::Mn_third_order_operator}.
	The equivalence of the results verified that~\eqref{Eq::Mn_Matrix_Exponents} is indeed the generating function of the Magnus terms.  
	They read as 
	\begin{align}
		\mathbf{M}_n^{[\ell]}(t,t_0)
		&= w_n^\ell [t-t_0]^\ell f_\ell\left(\frac{\Delta\omega[t-t_0]}{2}\right)
	\notag\\
		&\times\begin{pmatrix}
			0				&	e^{- i \Delta\omega [t+t_0]/2}	\\
			e^{ i \Delta\omega [r+r_0]/2}	&	0
		\end{pmatrix}
	\end{align}
	for $\ell$ odd and 
	\begin{align}
		\mathbf{M}_n^{[\ell]}(t,t_0)
		&=i w_n^\ell [t-t_0]^\ell f_\ell\left(\frac{\Delta\omega[t-t_0]}{2}\right)
		\begin{pmatrix}
			1				&	0	\\
			0	&			-1
		\end{pmatrix}
	\end{align}
	for $\ell$ even, 
	with the functions
	\begin{align}
		&f_1(z)=j_0(z)\text{,}\\
		&f_2(z)=\frac{1}{2}\left[j_1(z) \cos (z)-j_0(z) \sin (z)\right]\text{,}\\
		&f_3(z)=\frac{1}{6}\left[-j^3_0(z)+j_0(z)+j_2(z)\right]\text{,}\\
		&f_4(z)=\frac{1}{12} \left[\frac{1}{2} j_0^2(z) \sin (2 z)-\frac{1}{2} j_0(z) \sin (z)\right.
	\notag\\
		&-\frac{1}{2} j_1^2(z) \sin (2 z)-\frac{1}{2} j_2(z) \sin (z)-j_1(z) j_0(z) \cos (2 z)
	\notag\\	
		&\left.+\frac{3}{10} j_1(z) \cos (z)+\frac{3}{10} j_3(z) \cos (z)\right]
		\text{,}
	\\
 		&f_5(z)=\frac{1}{60} \left[\frac{j_0(z)}{2}+\frac{5 j_2(z)}{7}+\frac{3 j_4(z)}{14}\right. 
	\notag\\
 		&+2 j_1^2(z) j_0(z) \sin ^2(z)-\frac{13}{6} j_1(z) j_0(z) \sin (z)
 	\notag\\	
 		&-\frac{1}{2} j_3(z) j_0(z) \sin (z)-\frac{5}{3} j_1(z) j_2(z) \sin (z)
 	\notag\\
		&+2 j_0^3(z) \cos ^2(z)-\frac{5}{2} j_0^2(z) \cos (z)
 	\notag\\
 		&\left.-\frac{5}{2} j_2(z) j_0(z) \cos (z)+4 j_1(z) j_0^2(z) \sin (z) \cos (z)\right]
	\end{align}
	that are defined in terms of the pole free spherical Bessel functions $j_0(z)=\mathrm{sinc}(z)$.
	Evaluating the corresponding matrix exponentials, e.g., in second order 
	\begin{align}
		\mathbf{U}_n^{[2]}(t,t_0)=e^{-i|\kappa| \mathbf{M}_n^{[1]}(t,t_0)-|\kappa|^2 \mathbf{M}_n^{[2]}(t,t_0)}
	\end{align}
	yields the corresponding approximations for $a_n(t,t_0)$ and $b_n(t,t_0)$, i.e., $a_n^{[\ell]}(t,t_0)$ and $b_n^{[\ell]}(t,t_0)$.

\section{Convergence treatment}\label{App::Convergence_Analysis}

	The replacement $\tau_n\mapsto z$ in the matrix elements of $\mathbf{\tilde M }_\Lambda(\tau_n)$---cf.~\eqref{Eq::aLambda_bLambda_MLambda}---is performed after the conjugations in~\eqref{Eq::Mn_Matrix_Exponents}, i.e., $z$
	itself is not conjugated.
	In this manner we get the representations as 
	\begin{align}
		{\mathbf {\tilde M}}_\Lambda(z)=&\frac{\mathrm{arccos}(A_{R,\Lambda}(z))}{\sqrt{1-A_{R,\Lambda}^2(z)}}
	\notag\\
		&\times
		\begin{pmatrix}
			-A_{I,\Lambda}(z)	&	 e^{-i \Delta\omega [t+t_0]/2}B_\Lambda(z)	\\
			 e^{i \Delta\omega [t+t_0]/2}B_\Lambda(z)			&	A_{I,\Lambda}(z)
		\end{pmatrix}
		\text{,}
		\label{Eq::Mn_Matrix_Exponents_Replaced_kappa_by_z}
	\end{align}
	with
	\begin{align}
		&A_{R,\Lambda}(z)=\cos\left(\Lambda\right)\cos\left(\gamma_\Lambda(z)\right)+\Lambda \sin\left(\Lambda\right)\mathrm{sinc}\left(\gamma_\Lambda(z)\right)\text{,}
	\notag\\
		&A_{I,\Lambda}(z)=-\sin\left(\Lambda\right)\cos\left(\gamma_\Lambda(z)\right)+\Lambda \cos\left(\Lambda\right)\mathrm{sinc}\left(\gamma_\Lambda(z)\right)\text{,}
	\notag\\
		&\text{and }
		B_n(z)=z \,
		\mathrm{sinc}\left(\gamma_n(z)\right)
		\label{Eq::AI_AR_B}
	\end{align}
	where $\gamma_\Lambda(z)=\sqrt{\Lambda^2+  z^2}$.
	Replacing $z\mapsto|\kappa|$ in~\eqref{Eq::Mn_Matrix_Exponents_Replaced_kappa_by_z} yields~\eqref{Eq::Mn_Matrix_Exponents}. Thus, \eqref{Eq::Mn_Matrix_Exponents_Replaced_kappa_by_z} is a  continuation of~\eqref{Eq::Mn_Matrix_Exponents}.  

	First, let us note that there is no branching in the functions~\eqref{Eq::AI_AR_B} as 
	the square root $\gamma_\Lambda(z)$ only appears in the even $\cos$- and $\mathrm{sinc}$-functions.
	Furthermore, let us note that with the generating function~\cite{AbramowitzS10.1.40}
	\begin{align}
		\frac{1}{Z}\cos\left(\sqrt{Z^2-2ZT}\right)=\sum\limits_{p=0}^{\infty}  \frac{T^p}{p!}j_{p-1}(Z)
	\end{align}
	of the spherical Bessel functions 
	\begin{align}
		j_{-1}(Z)&=\frac{\cos(Z)}{Z}
	\notag\\
		j_p(Z)&=(-Z)^p\left(\frac{1}{Z}\frac{d}{d Z}\right)^p \frac{\sin(Z)}{Z} \text{ for }p=0,1,\dots
	\end{align}
	and its derivative in terms of $T$ we can find the Maclaurin series representations
	\begin{align}
		\cos\left(\gamma_\Lambda(z)\right)&=\Lambda\sum\limits_{p=0}^{\infty}  \frac{1}{p!} \left(\frac{-z^2}{2\Lambda} \right)^p j_{p-1}(\Lambda)
	\notag\\
		\mathrm{sinc}\left(\gamma_\Lambda(z)\right)&=\sum\limits_{p=0}^{\infty}  \frac{1}{p!}\left(\frac{-z^2}{2\Lambda} \right)^p j_{p}(\Lambda)
		\text{.}
	\end{align}
	Note that the series are entirely independent of the conjugated complex variable $z^*$.
	
	With help of $|j_p(Z)|\leq 1$ for $p=0,1,\dots$ and $Z\in[0,\infty) $ we can show absolute convergence of these series
	with upper bounds
	\begin{align}
		|\cos\left(\gamma_\Lambda(z)\right)|&\leq |\cos(\Lambda)|+
		\Lambda \left(\exp\left[
		\frac{|z|^2}{2|\Lambda|} 
		\right]-1\right)
	\notag\\
		\text{and }
		|\mathrm{sinc}\left(\gamma(z)\right)|&\leq 
		\exp\left[
		\frac{|z|^2}{2|\Lambda|} 
		\right]
		\text{.}
	\end{align}
	Thus, the functions $A_{I,\Lambda}(z)$, $A_{R,\Lambda}(z)$, and $B_{\Lambda}(z)$ defined in~\eqref{Eq::AI_AR_B} are analytical functions in the full complex plane $z\in\mathbb{C}$.
	Thus, singularities of $\mathbf{\tilde M}_\Lambda(z)$ can only stem from the factor
	\begin{align}
		f(A_{R,\Lambda}(z))=\frac{\mathrm{arccos}(A_{R,\Lambda}(z))}{\sqrt{1-A_{R,\Lambda}^2(z)}}
		\text{.}
		\label{Eq::A_R}
	\end{align}

	Note that  
	\begin{align}
		f(z)=\frac{dF(z)}{dz} \text{ with }F(z)=-\frac{1}{2}\mathrm{arccos}^2(z)
		\text{.}
	\end{align}
	One can show, that the function $F(z)$ has a branch point at $z=-1$   (but unlike  $\mathrm{arccos}(z)$ not at $z=1$)--a beautiful introduction to the concepts of branch points and branch cuts can be found in Ref.~\cite{Needham1997}. Consequently $f(z)$ has the same branch point as $F(z)$. 
	Thus, the function $f(A_{R,\Lambda}(z))$ has  branch points wherever $A_{R,\Lambda}(z)=-1$, i.e.,
	the branch points of $f(A_{R,\Lambda}(z))$  correspond to the zeros of the analytic function
	\begin{align}
		g_\Lambda(z)=A_{R,\Lambda}(z)+1
		\text{.}
	\end{align}
	As we have shown that all other functions appearing in~\eqref{Eq::Mn_Matrix_Exponents_Replaced_kappa_by_z} are analytic, $\mathbf{\tilde M}_\Lambda(z)$ also has branch points, where $g_\Lambda(z)=0$.
	The branch cut lines---originating from the branch points---can always be chosen such that they point away from the origin and do not cross. 
	Thus, a series expansion of $\mathbf{\tilde M}_\Lambda(z)$ around $z=0$ will converge for $|z|< r_\Lambda $  where 
	\begin{align}
		\label{Eq:resultAppB}
		 r_\Lambda= \min\limits_{z_0\in \mathbb{C}:g_\Lambda(z_0)=0} |z_0|
		 \text{.}
	\end{align}
	
	We have evaluated $r_\Lambda$ in a range of $\Lambda$ going from $\Lambda=0.005 \pi$ to $\Lambda =200\pi$ in steps of $0.005\pi$. This was achieved by extracting the line data from the ContourPlot function (Contours: $\mathrm{Re}(g_\Lambda(z))=0$, $\mathrm{Im}(g_\Lambda(z))=0$) in Mathematica to get estimates for the location of the minimal absolute value zeros of $g(z)$ which where then refined by the FindRoot function in Mathematica.

	
\end{document}